\documentclass[aps,prx,showpacs,superscriptaddress,nofootinbib,notitlepage,twocolumn]{revtex4-1}
\usepackage{graphicx}
\usepackage{color}
\usepackage{transparent}
\usepackage{amsmath,amssymb,amsfonts,amssymb}
\usepackage{epstopdf}
\usepackage{float}
\usepackage{hyperref}

\newcommand{\bra}[1]{\ensuremath{\left\langle#1\right|}}
\newcommand{\ket}[1]{\ensuremath{\left|#1\right\rangle}}

\newcommand{\tr}[0]{\text{tr}}

\newcommand{\mc}[1]{\mathcal{#1}} 
\newcommand{\sh}{\mathcal{S}(\mathcal{H})} 
\newcommand{\hi}{\mathcal{H}} 

\begin{document}

\title[]{Experimental Quantum Probing Measurements With No Knowledge on the System-Probe Interaction}

\author{Henri Lyyra}
\affiliation{Laboratory of Quantum Optics, Department of Physics and Astronomy, University of Turku, FI-20014, Turun yliopisto, Finland}
\affiliation{Turku Centre for Quantum Physics, Department of Physics and Astronomy, University of Turku, FI-20014, Turun yliopisto, Finland}
\affiliation{QTF Centre of Excellence, Department of Physics and Astronomy, University of Turku, FI-20014 Turun Yliopisto, Finland}
\affiliation{Department of Physics and Nanoscience Center, University of Jyv\"askyl\"a, FI-40014 University of Jyv\"askyl\"a, Finland}
\email{henri.s.lyyra@jyu.fi}

\author{Olli Siltanen}
\affiliation{Laboratory of Quantum Optics, Department of Physics and Astronomy, University of Turku, FI-20014, Turun yliopisto, Finland}
\affiliation{Turku Centre for Quantum Physics, Department of Physics and Astronomy, University of Turku, FI-20014, Turun yliopisto, Finland}
\affiliation{QTF Centre of Excellence, Department of Physics and Astronomy, University of Turku, FI-20014 Turun Yliopisto, Finland}

\author{Jyrki Piilo}
\affiliation{Laboratory of Quantum Optics, Department of Physics and Astronomy, University of Turku, FI-20014, Turun yliopisto, Finland}
\affiliation{Turku Centre for Quantum Physics, Department of Physics and Astronomy, University of Turku, FI-20014, Turun yliopisto, Finland}
\affiliation{QTF Centre of Excellence, Department of Physics and Astronomy, University of Turku, FI-20014 Turun Yliopisto, Finland}

\author{Subhashish Banerjee}
\affiliation{Indian Institute of Technology Jodhpur, Jodhpur 342011, India}

\author{Tom Kuusela}
\affiliation{Laboratory of Quantum Optics, Department of Physics and Astronomy, University of Turku, FI-20014, Turun yliopisto, Finland}
\affiliation{Turku Centre for Quantum Physics, Department of Physics and Astronomy, University of Turku, FI-20014, Turun yliopisto, Finland}



\begin{abstract}

In any natural science, measurements are the essential link between theory and observable reality. 
Is it possible to obtain accurate and relevant information via measurement whose action on the probed system is unknown? In other words, can one be convinced to know something about the nature without knowing in detail how the information was obtained? In this paper, we show that the answer is surprisingly, yes. We construct and experimentally implement a quantum optical probing measurement where measurements on the probes, the photons' polarization states,  are used to extract information on the systems, the frequency spectra of the same photons. Unlike the pre-existing probing protocols, our measurement does not require any knowledge of the interaction between the probe and the system.
\end{abstract}

\maketitle

\section{Introduction}

As a necessary part of any natural science, measurements lie in the heart of understanding the observable universe. They are needed both to test the existing theories and to inspire new branches of theoretical research. In addition to purely scientific purposes, measurements are necessary also for technological applications. More specifically, the readout of the outcome of quantum computer or quantum simulator is extracted by measuring the system acting as the information carrier.  Sometimes, measurements can be needed to also monitor the performance of the device while it is running.

Despite their useful purposes, measurements have also harmful effects. Every non-trivial measurement disturbs the system state \cite{bennett,fuchs,maccone}. In some cases, direct measurement may even destroy the whole system of interest. To avoid the problem, so-called non-demolition \cite{probingHaroche,probingRaha} and quantum probing measurements have been proposed and experimentally implemented \cite{probing2006,probingnonmark2013,probing2013,probing2016,probing2017}. In quantum probing measurement, the idea is to avoid directly measuring the system of interest by coupling it to a disposable \emph{probe system} instead. The system-probe coupling causes the probe state to change and from the change one can deduce how the system of interest was before the coupling \cite{petru,RHbook,banerjeebook}. The existing protocols are heavily based on fully knowing the system-probe coupling and implemententation of such protocols requires faithful experimental realization of that specific coupling. As any experimental implementation has its limitations, the coupling is never perfectly known nor controlled.

In addition to the above mentioned quantum probing schemes, it is also worth mentioning here recent developments in quantum hypothesis testing, see, e.g.,~\cite{molmer-ht,tan,wilde-ht}. The motivating question -- for hypothesis testing -- is which one of the {\it{a priori}} introduced hypotheses is consistent with the obtained measurement data. For example in the open system context, it is possible to test which one of the possible Rabi frequencies was actually used to drive a two-level
 system~\cite{molmer-ht}. Moreover, hypothesis testing has also applications, e.g., in quantum illumination \cite{pirandola} to check whether a low-reflectivity object existed -- or not -- in a given target region~\cite{tan,wilde-ht}. For quantum probing, the starting point is slightly different. Here, the aim is,
 in the best case, to obtain precise quantitative value --  with bounds and 
 without {\it {prior}} information or hypotheses --  on a property of a given degree of freedom whilst measuring another degree of freedom.

In \cite{tukiainen}, a new approach to quantum probing was proposed. These protocols are based on the properties of so-called $\alpha$-fidelities, which were shown to satisfy a generalized data processing inequality which was found useful for multiple purposes \cite{tukiainen}. The inequality was applied to study the Hilbert space dimension of the programmable quantum processor in approximate implementation of quantum channels, and making predictions of unsolvable quantum dynamics. Interestingly, it was also shown to allow for constructing probing measurements without knowing anything about the system-probe coupling. In this sense, the protocol should give accurate information even though it is impossible to know how the measurement actually happens. 

In this paper, we present the first experimental implementation of such probing measurement. Our system of interest is the frequency degree of freedom of a single photon and our probe is the polarization of the same photon. This system has raised a lot of attention lately \cite{liu2011,setuphamiltonian,karlssonLyyra,karlssonEPL,sina2017,liu,sina2020,olli,hefeiTele}. We show how measuring two polarization states before and after their interaction with the frequency can be used to extract upper bounds for the width of the corresponding frequency spectra. We demonstrate how the protocol works even in cases where the coupling is unknown. As a trade-off of performing the probing without knowing the system-probe coupling, we need to perform full tomography for the probe system. Our measurement data does not yield to estimate for the exact value of the unknown parameter, but instead  we obtain upper or lower bounds, derived analytically from the generalized data processing inequality.

The paper is structured as follows: In Sec.~\ref{theory}, we briefly discuss the open quantum system picture, the $\alpha$-fidelities and the generalized data processing inequality. In Sec.~\ref{photonicprobe}, we present our photonic system and apply the generalized data processing inequality to derive the bounds for the unknown width to be determined by the probing measurement. In Sec.~\ref{experiment}, we present the experimental setup and the measurement results. Finally, in Sec.~\ref{conclusions} we summarize our results and discuss future aspects.

\section{Open Quantum Systems, Quantum Probing, and $\alpha$-fidelities}\label{theory}

We say that a quantum system $A$ is open if it interacts with some other system $B$, the environment. Commonly, it is assumed that $A$ and $B$ are uncorrelated before the dynamics begins. In the dynamics, the total state of the combined system $AB$  undergoes a change, described by a unitary $U$. The evolved state of system $A$ can be solved as
\begin{equation}\label{reduced}
\Phi(\rho) = \text{tr}_B[ U ( \rho \otimes \xi ) U^\dagger ]\,,
\end{equation}
where $\rho$ and $\xi$ are the initial states of systems $A$ and $B$, respectively, and tr$_B[X]$ is the partial trace of $X$ over the Hilbert space of $B$ \cite{petru,RHbook,banerjeebook}. By this construction, $\Phi$ is a completely positive and trace preserving (CPTP) map, or in other words a \emph{channel}. The effects of CPTP maps have been widely studied and it has been shown that information in terms of trace distance \cite{nielsenBook} and quantum entanglement can only be lost in (local) CPTP transformations \cite{bennettEntanglement}, and fidelity between two states can never decrease \cite{nielsenBook}. For trace distance and fidelity, this means that they satisfy data processing inequalities.

\begin{figure}[t]
\centering
  \includegraphics[width=0.95\linewidth]{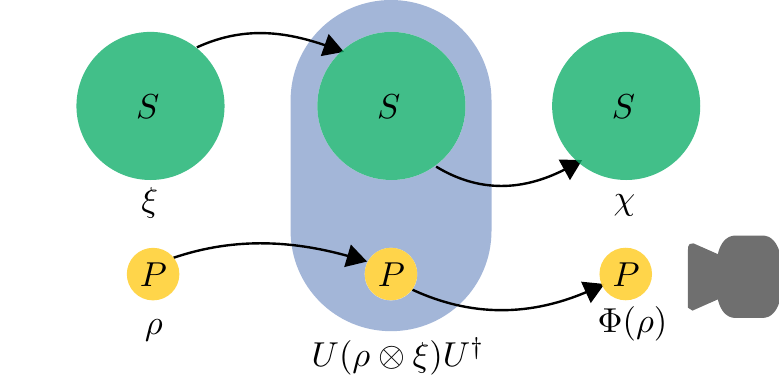}
  \caption{The common quantum probing approach (color online). The system $S$ and the probe $P$ interact under the unitary coupling $U$. After the interaction, measurements on $P$ in the evolved state $\Phi(\rho)$ are used to determine unknown properties of $S$. Here $U$ is known and the unknown parameters of $\xi$ are mapped to the state transformation $\rho \mapsto \Phi(\rho)$ of $P$.}
  \label{env_in_dyn}
\end{figure}

Despite its harmful effects, open system dynamics can also be useful. One of its applications is the \emph{quantum probing} measurements. In quantum probing, the goal is to obtain information of some property of the system $S$ without directly measuring it. This can be the case when $S$ is a part of a device, such as a quantum computer or a quantum simulator, and one wants to monitor the device without having to stop it to perform a measurement. In quantum probing, $S$ is unitarily coupled to a disposable probe system $P$ and measurements on the evolved probe are used to gain information about $S$. In the above description of open quantum system dynamics, the system of interest $S$ corresponds to the environment $B$ and the probe $P$ is the open system $A$. Commonly, the probing protocols rely on knowing the coupling $U$ and they are based on the solvable connection between the unknown parameters of $S$ and the channel $\Phi$ of $P$, caused by the interaction, as in Fig.~\ref{env_in_dyn}.

\begin{figure}[t]
\centering
  \includegraphics[width=0.95\linewidth]{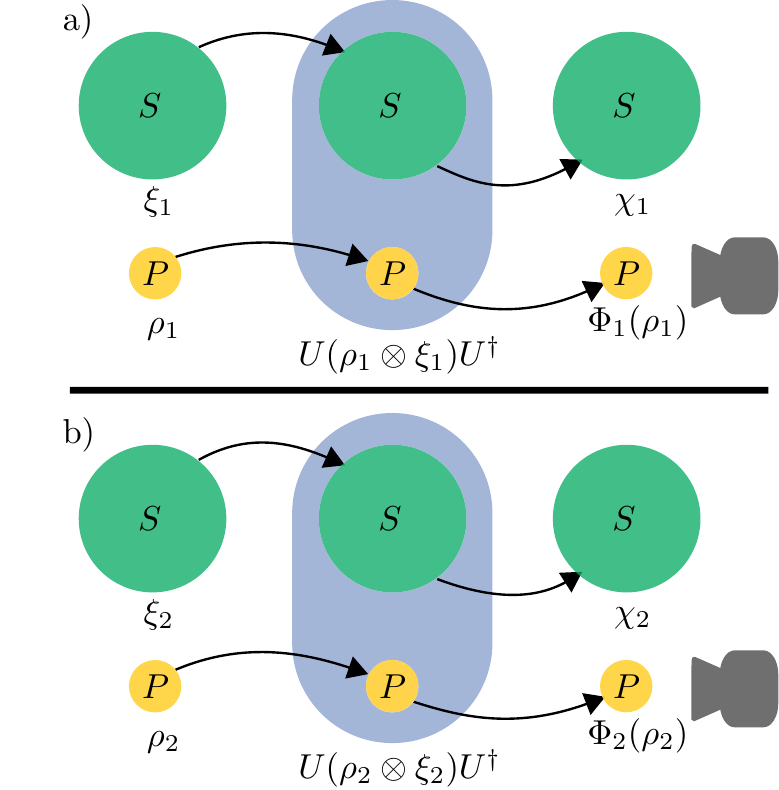}
  \caption{The unknown coupling quantum probing approach (color online). The system $S$ interacts with the probe $P$. In cases a) and b), the unitary coupling $U$ between $S$ and $P$ is the same. The unknown parameter in the system states $\xi_1$ and $\xi_2$ is the same, but some control parameter is different. As a consequence, the induced probe channels $\Phi_1$ and $\Phi_2$ may be different. Even though the coupling $U$ is unknown, and consequently so are the channels $\Phi_1$ and $\Phi_2$, comparison of the measured probe states $\Phi_1(\rho_1)$ and  $\Phi_2(\rho_2)$ can be used to gain accurate information on the unknown parameter.}
  \label{env_in_dyn-12}
\end{figure}

On the other hand if the coupling $U$ is not known, the unknown parameters of $S$ cannot be mapped to the transformation $\Phi$ of the probe state. Consequently, the traditional probing approach cannot be applied. To see how the quantum probing can be performed in such situation, let us consider the two cases in Fig.~\ref{env_in_dyn-12}. When the coupling $U$ has been fixed, the dynamics of the probe $P$ depends on the initial state $\xi$ of the system $S$. As a consequence, preparing $S$ in different states $\xi_1$ and $\xi_2$ and coupling it to $P$ can induce different channels $\Phi_1$ and $\Phi_2$ to $P$ even if the coupling $U$ is the same in both cases. 

This observation was exploited in \cite{tukiainen} to form a mathematical tool for studying open quantum systems based on the comparison between the initial environment states and the channels they induce. The {\it $\alpha$-fidelity of states} was defined for $\alpha \in (0,1)$ as \cite{tukiainen}
\begin{equation}
F_{\alpha} \big(\rho_1, \rho_2\big) := \tr \left[ \left(\rho_2^{\frac{1-\alpha}{2\alpha}} \rho_1 \, \rho_2^{\frac{1-\alpha}{2\alpha}} \right)^\alpha \right]\,.
\end{equation}
In the special case $\alpha=1/2$, we note that $F_{1/2}$ is the commonly used fidelity of states.

Now, let us consider the $\alpha$-fidelities in the context of Fig.~\ref{env_in_dyn-12}. The unitary coupling $U$ between $P$ and $S$ is fixed but in \ref{env_in_dyn-12} a) and b) the initial states of $P$ and $S$ can be different. Thus, different choices of states $\xi_1$ and $\xi_2$ of $S$ induce channels $\Phi_1$ and $\Phi_2$ to $P$ in the interaction, respectively. 
In this open system picture, it was shown that 
the $\alpha$-fidelities satisfy the following inequality
\begin{equation}\label{eq:mainineq2}
F_{\alpha} \big(\rho_1, \rho_2\big) F_{\alpha} \big(\xi_1, \xi_2\big) \leq F_{\alpha} \big(\Phi_1(\rho_1), \Phi_2(\rho_2)\big)
\end{equation} 
for all $\alpha\in[1/2,1)$ \cite{tukiainen}. It is worth noting that Eq.~\eqref{eq:mainineq2} does not explicitly depend on the coupling $U$, so it gives a fundamental bound for the relation of the channels $\Phi_1$ and $\Phi_2$ that two states $\xi_1$ and $\xi_2$ can induce. Conversely, it sets restrictions to the states $\xi_1$ and $\xi_2$ that can induce two given channels $\Phi_1$ and $\Phi_2$. This property makes it useful for different applications. Equation \eqref{eq:mainineq2} can also be interpreted as a generalization of the data processing inequality $F_{1/2} \big(\rho_1, \rho_2\big) \leq F_{1/2} \big(\Phi(\rho_1), \Phi(\rho_2)\big)$ of the common fidelity function by adding the freedom of parameter for $\alpha\in[1/2,1)$, giving us $F_{\alpha} \big(\rho_1, \rho_2\big) \leq F_{\alpha} \big(\Phi(\rho_1), \Phi(\rho_2)\big)$, and even further to the case of different channels $\Phi_1$ and $\Phi_2$, as in Eq.~\eqref{eq:mainineq2}.  

Interestingly, Eq.~\eqref{eq:mainineq2} allows us to construct quantum probing protocols with no knowledge of the coupling in the following way: Assume that $S$ is prepared in the state $\xi(x,y)$ and our task is to extract information of the value of parameter $x$, and $y$ is some controllable parameter. The experimenter has control over the initial state of $P$ and the parameter $y$. By preparing $P$ in some known states $\rho_1$ and $\rho_2$, and evolving them with the channels $\Phi_1$ and $\Phi_2$, induced by states $\xi(x,y_1)$ and $\xi(x,y_2)$ of $S$, respectively, the experimenter obtains the values of $F_{\alpha} \big(\rho_1, \rho_2\big)$ and $F_{\alpha} \big(\Phi_1(\rho_1), \Phi_2(\rho_2)\big)$. When the $x$ and $y$ dependence of $\xi(x,y)$ is known, different values of $x$ can be numerically tested in $F_{\alpha} \big(\xi(x,y_1), \xi(x,y_2)\big)$. The values of $x$ which cause violation of  Eq.~\eqref{eq:mainineq2} are immediately known to be incorrect and bounds of the actual value of $x$ can be obtained. In cases where $F_{\alpha} \big(\xi(x,y_1), \xi(x,y_2)\big)$ is bijective in terms of $x$, analytical bounds for the unknown $x$ can be derived as functions of $F_{\alpha} \big(\rho_1, \rho_2\big)$, $F_{\alpha} \big(\Phi_1(\rho_1), \Phi_2(\rho_2)\big)$, $y_1$, and $y_2$.

Next, we construct and analyze such a protocol by fixing the system and probe, and then we implement the probing protocol in an all-optical experiment.

\section{The Photonic system and probing}\label{photonicprobe}

\begin{figure}[t]
\centering
  \includegraphics[width=0.95\linewidth]{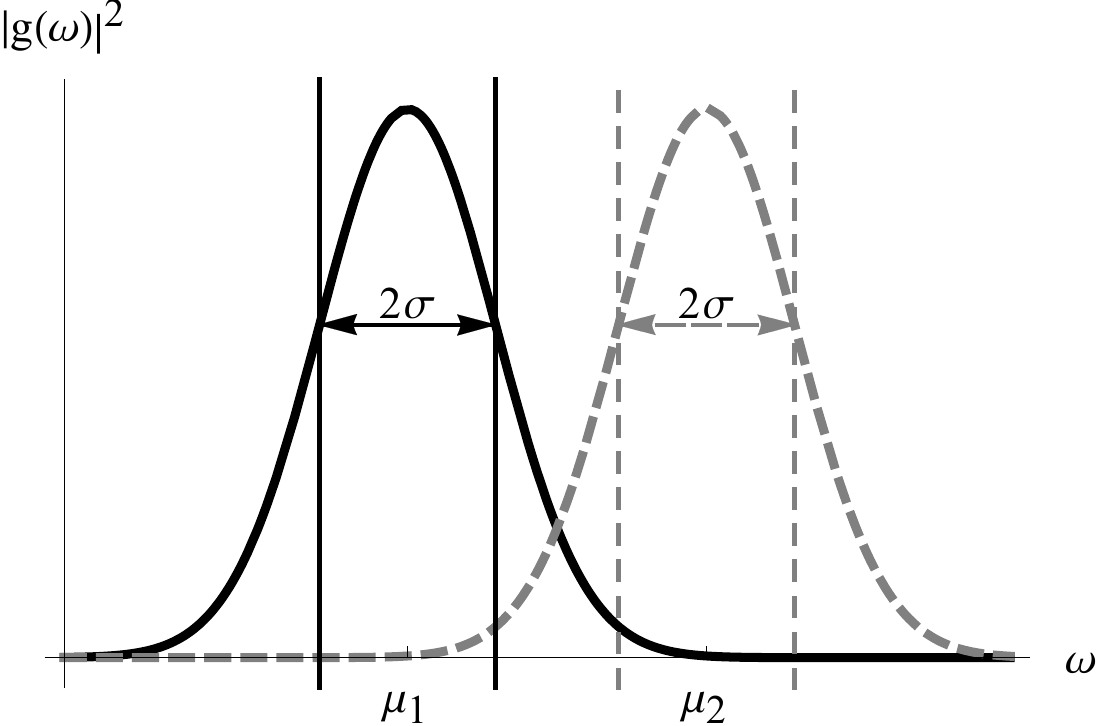}
  \caption{
Illustration of the frequency states $\xi_1$ and $\xi_2$. The standard deviation $\sigma$ is the same in both Gaussian distributions $| g_1(\omega) |^2$ and $| g_2(\omega) |^2$, and it is the unknown parameter of our interest. In the probing protocol, we control the difference between central frequencies $\Delta\mu = | \mu_2 - \mu_1 |$, and thus it is known.
  }
  \label{gaussians}
\end{figure}

Our system of interest is the frequency degree of freedom of a single photon. We assume, that the central frequency $\mu$ of Gaussian intensity distribution $|g( \omega )|^2$ can be shifted in a controlled way. Instead, the standard deviation $\sigma$ is unknown, and our goal is to get information of it. The frequency states are given by 
\begin{align}\label{systemStates}
\xi_k &= \int |g_k( \omega )|^2 \ket{\omega} \bra{\omega} d \omega \,,  \text{ where} \\
| g_k(\omega) |^2 &= \frac{1}{ \sqrt{2 \pi \sigma^2} } e^{- \frac{ (\omega - \mu_k )^2}{2 \sigma^2}}\,,~k\in \{1,2\} \,, 
\end{align}
$\sigma$ is the unknown standard deviation, and $\mu_k$ is the mean or central frequency of the Gaussian distribution $| g_k(\omega) |^2$, as illustrated in Fig~\ref{gaussians}. 
Here,  $\omega$ are the frequency values with amplitudes  $g_k(\omega)$. 
Now, the $\alpha$-fidelity between two frequency states becomes
\begin{equation}\label{aFidInitSolution}
F_\alpha (\xi_1, \xi_2) = e^{- (1 - \alpha ) \alpha \frac{ \Delta \mu^2 }{ 2 \sigma^2 } }\,,
\end{equation}
where we have denoted $\Delta\mu = | \mu_2 - \mu_1 |$. In the experiment, $\Delta\mu$ is our known controllable parameter. We note that $F_\alpha (\xi_1, \xi_2)$ is monotonically increasing in $\sigma$ and $\alpha$, and monotonically decreasing in $\Delta\mu$ when $\alpha \in [1/2, 1)$. 
%
In this optical setup, our probe is the polarization qubit of the photon.
 
Now, we consider what kinds of bounds can be derived from Eq.~\eqref{eq:mainineq2} for the three cases $F_\alpha ( \Phi_1 (\rho_1), \Phi_2 (\rho_2) ) < F_\alpha ( \rho_1, \rho_2)$, $F_\alpha ( \Phi_1 (\rho_1), \Phi_2 (\rho_2) ) = F_\alpha ( \rho_1, \rho_2)$, and $F_\alpha ( \Phi_1 (\rho_1), \Phi_2 (\rho_2) ) > F_\alpha ( \rho_1, \rho_2)$. First, assuming $F_\alpha \big( \Phi_1 (\rho_1), \Phi_2 (\rho_2) \big) < F_\alpha ( \rho_1, \rho_2 )$, $\Delta\mu > 0$, and using the initial system states of Eq.~\eqref{systemStates}, equation \eqref{eq:mainineq2} yields to
\begin{equation}\label{ourbound1}
\sigma \leq B_1(\alpha) :=\sqrt{ \frac{ \alpha( \alpha - 1 )  \Delta\mu^2 }{ 2 \log \big[ F_\alpha \big( \Phi_1(\rho_1), \Phi_2(\rho_2) \big) / F_\alpha ( \rho_1, \rho_2 ) \big] } } 
\,.
\end{equation}
Thus, measuring $\Phi_1(\rho_1)$ and $\Phi_2(\rho_2)$ results directly to an $\alpha$ parametrized family of upper bounds for the unknown standard deviation $\sigma$. We note that in our case $F_\alpha (\xi_1,\xi_2) = F_\alpha (\xi_2,\xi_1)$, but generally $F_\alpha$ is not symmetric w.r.t.~the inputs, namely $F_\alpha \big( \Phi_1(\rho_1),\Phi_2(\rho_2) \big) \neq F_\alpha \big( \Phi_2(\rho_2),\Phi_1(\rho_1) \big)$ and $F_\alpha ( \rho_1, \rho_2 ) \neq F_\alpha ( \rho_2, \rho_1 )$. As a consequence, the same measurement data gives us an additional family of bounds:
\begin{equation}\label{ourbound2}
\sigma \leq B_2(\alpha) := \sqrt{ \frac{ \alpha( \alpha - 1 )  \Delta\mu^2 }{ 2 \log \big[ F_\alpha \big( \Phi_2(\rho_2), \Phi_1(\rho_1)  \big)  / F_\alpha ( \rho_2, \rho_1 ) \big] } } 
\,.
\end{equation} 

On the other hand, if $F_\alpha \big( \Phi_1 (\rho_1), \Phi_2 (\rho_2) \big) = F_\alpha ( \rho_1, \rho_2 )$ or $\Delta\mu = 0$, the initial system states of Eq.~\eqref{systemStates} in Eq.~\eqref{eq:mainineq2} lead to the trivial condition $\sigma \ge 0$. 
Lastly, assuming $F_\alpha ( \rho_1, \rho_2 ) < F_\alpha \big( \Phi_1 (\rho_1), \Phi_2 (\rho_2) \big)$, $\Delta\mu > 0$, and using the initial system states of Eq.~\eqref{systemStates} in Eq.~\eqref{eq:mainineq2} gives us
\begin{align}\label{ourbound1b}
\sigma^2 &\geq  \frac{ \alpha( \alpha - 1 )  \Delta\mu^2 }{ 2 \log \big[ F_\alpha \big( \Phi_1(\rho_1), \Phi_2(\rho_2) \big) / F_\alpha ( \rho_1, \rho_2 ) \big] } 
\,,
~\text{and} \\
\label{ourbound2b}
\sigma^2 &\geq   \frac{ \alpha( \alpha - 1 )  \Delta\mu^2 }{ 2 \log \big[ F_\alpha \big( \Phi_2(\rho_2), \Phi_1(\rho_1)  \big)  / F_\alpha ( \rho_2, \rho_1 ) \big] }  
\,,
\end{align} 
as above. With the above assumptions, the right-hand side in Eq.~\eqref{ourbound1b} and \eqref{ourbound2b} is genuinely negative. Thus, if $F_\alpha ( \rho_1, \rho_2 ) < F_\alpha \big( \Phi_1 (\rho_1), \Phi_2 (\rho_2) \big)$, the probing protocol does not give us any relevant information. This observation suggests that in order to guarantee relevant information, one should maximize  $F_\alpha ( \rho_1, \rho_2 )$ by choosing $\rho_1 = \rho_2 = \rho$, as there is very little control over $F_\alpha \big( \Phi_1 (\rho_1), \Phi_2 (\rho_2) \big)$ in the case of unknown coupling $U$. 

We note here, that due to the similar role of $\sigma$ and $\Delta\mu$ in Eq.~\eqref{aFidInitSolution}, our probing protocol can be used to get bounds for unknown frequency shift $\Delta\mu$ if $\sigma$ was known instead. In this case, the same measurement data could be used and lower bounds for $\Delta\mu$ would be determined as
\begin{align}
\Delta\mu & \ge \sqrt{ \frac{ 2\sigma^2 \log \big[ F_\alpha \big( \Phi_1(\rho_1), \Phi_2(\rho_2)  \big)  / F_\alpha ( \rho_1, \rho_2 ) \big] }{ \alpha( \alpha - 1 ) } }\,, \\
\Delta\mu & \ge \sqrt{ \frac{ 2\sigma^2 \log \big[ F_\alpha \big( \Phi_2(\rho_2), \Phi_1(\rho_1)  \big)  / F_\alpha ( \rho_2, \rho_1 ) \big] }{ \alpha( \alpha - 1 ) } }\,,
\end{align}
when $F_\alpha \big( \Phi_1(\rho_1), \Phi_2(\rho_2)  \big) < F_\alpha ( \rho_1, \rho_2 )$ and $F_\alpha \big( \Phi_2(\rho_2), \Phi_1(\rho_1)  \big) < F_\alpha ( \rho_2, \rho_1 )$, respectively. Similarly, if both $\Delta\mu$ and $\sigma$ were unknown, we could get lower bounds for their ratio $\Delta\mu / \sigma$. 

The above analysis was performed for the full generality of the $\alpha$ fidelities but the same would hold also for the common fidelity of quantum states, obtained by choosing $\alpha = 1/2$. As suggested by the theoretical results in \cite{tukiainen}, freedom to choose $\alpha$ can lead to improved precision in probing protocols, so we exploit here the whole range $\alpha \in [1/2,1)$ to get as tight bounds as possible.

Next, we implement the two frequency states with $\Delta\mu > 0$ and experimentally determine the values of $F_\alpha ( \rho_1, \rho_2 )$, $F_\alpha ( \rho_2, \rho_1 )$, $F_\alpha \big( \Phi_1 (\rho_1), \Phi_2 (\rho_2) \big)$, and $F_\alpha \big( \Phi_2 (\rho_2), \Phi_1 (\rho_1) \big)$ to get upper bounds for $\sigma$ with our probing protocol.

\section{The Experiment}\label{experiment}

\begin{figure*}[t]
\begin{minipage}[t]{1\textwidth}
\centering
  \includegraphics[width=0.95\linewidth]{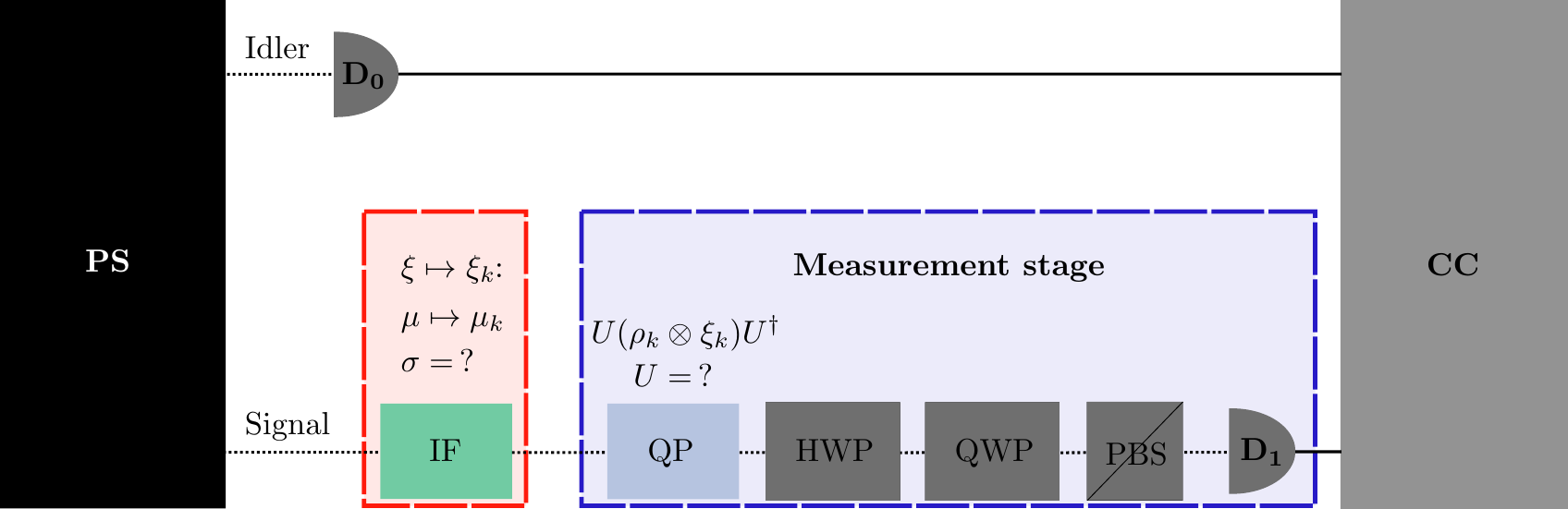}
  \caption{The experimental setup (color online). The photon source (PS) produces a pair of photons. The signal photon’s polarization state is prepared at PS to $\ket{+}\bra{+}$ by passing it through a polarizer fixed to 45 degree angle w.r.t.~the $\leftrightarrow$ axis. The idler photon is detected with the single photon detector D$_0$ which triggers the coincidence counting electronics (CC) to wait for the signal photon to arrive at detector D$_1$. After PS, the central frequency $\mu$ of the signal photon can be adjusted by 
  tilting the interference filter (IF). After the state preparation, the signal photon goes to the probing measurement stage. First, frequency (the system) is coupled with polarization (the probe) when the photon goes through quartz plates (QP). After the interaction, the evolved state $\Phi(\rho)$ of the polarization qubit (probe) is determined by performing a tomographic measurement with a combination of a half-wave plate (HWP), quarter-wave plate (QWP), and a polarizing beamsplitter (PBS).}
  \label{setup}
\end{minipage}
\end{figure*}

In figure \ref{setup}, we present the experimental setup. 
The photon source (PS) is a type-I beta-barium borate crystal, which is pumped with a tightly focused continuous wave laser of the wavelength 405 nm. The crystal produces randomly a pair of photons through the spontaneous parametric down-conversion process in a wide spectrum around 810 nm. At PS, \emph{the signal photon’s} polarization is prepared to an initial probe state $\rho$ by passing it through a polarizer rotated to a fixed angle w.r.t. the $\leftrightarrow$ plane. Here, $\updownarrow$ and $\leftrightarrow$ correspond to vertical and horizontal polarizations, respectively.
 \emph{The idler photon} in the upper branch is registered by the single photon detector D$_0$, which triggers the coincidence counting electronics (CC) to monitor the single photon detector D$_1$ for data collection of the signal photon in the lower branch. 

The signal photon goes first through the 
interference filter (IF). 
Tilting IF changes its transmission bandwidth and as a consequence, the central frequency $\mu$ of the Gaussian frequency distribution, while keeping its standard deviation $\sigma$ as it was. This controlled transformation allows us to change the initial frequency state $\xi$ into $\xi_1$ and $\xi_2$, and thus choose $\Delta\mu$ which needs to be non-zero for our protocol to work.

The probing measurement is performed at the measurement stage. First the system (frequency) and the probe (polarization) are coupled as the signal photon goes through birefringent quartz plates (QP)
. This causes the channels $\Phi_1$ and $\Phi_2$ which change the polarization states $\rho_1$ and $\rho_2$, respectively
\footnote{We emphasize that even though the dynamics induced by a QP coupling is well known once its optical axis is fixed, any assumption on the coupling or the channels is not used to derive our protocol and it would work similarly for any other coupling, as illustrated later by fixing randomly rotated orientations to each QP in combinations.}
. After the QP, the signal photon passes through a combination of a half-wave plate (HWP), a quarter-wave plate (QWP), and a polarizing beamsplitter (PBS). Rotating the QWP and the HWP changes the measurement basis of PBS, and allows for full state tomography of the polarization states $\Phi_1(\rho_1)$ and $\Phi_2(\rho_2)$. For each measurement basis, we used 60 s integration time.

Measuring how channels $\Phi_1$ and $\Phi_2$ change some initial probe states can give insight on how the optimal initial probe states should be chosen: In our case, polarization tomography after interaction with frequency shows that the diagonal terms in the $\{\leftrightarrow,\updownarrow\}$ basis remain constant and there is decay and rotation of the complex phase in the off-diagonal terms when the quartz plates in the experimental setup are fixed in the same orientation. This suggests dephasing type dynamics for the probe, for which optimal initial states were shown to be $\rho_1 = \rho_2 = \ket{+}\bra{+}$ where $\ket{+} = \frac{1}{\sqrt{2}} (\ket{\leftrightarrow} + \ket{\updownarrow})$ \cite{tukiainen}. As a consequence, we choose to prepare our initial probe states close to
\begin{align}
\rho_1 &= \rho_2 = \rho := \ket{+}\bra{+}\,.
\end{align}

The initial probe states in the experiment were determined to be
\begin{align}
\rho_1 &= 
\begin{pmatrix}
0.513	&	 0.482 - 0.006 i \\
0.482 + 0.06 i  	& 	0.487
\end{pmatrix}\,, \\
\rho_2 &= 
\begin{pmatrix}
0.535	&	 0.496 - 0.017 i \\
0.496 + 0.017 i  	& 	0.465
\end{pmatrix}\,,
\end{align}
where we have used the matrix representation $\ket{\leftrightarrow} = (1,\, 0)^\text{T}$ and $\ket{\updownarrow} = (0,\, 1)^\text{T}$. We fixed the control parameter as $\Delta\mu = 7.95 
\times 10^{11}$ Hz (in wavelength $\Delta\lambda = 1.73 
$ nm). We used multiple different thicknesses of quartz plates and their combinations, corresponding to different system-probe couplings $U$. As an example, in the case of a 5 mm quartz plate as coupling, the corresponding evolved probe states were
\begin{align}
\Phi_1 (\rho_1) &= 
\begin{pmatrix}
0.51	&	 0.435 + 0.073 i \\
0.435 - 0.073 i  	& 	0.49
\end{pmatrix}\,,\\
\Phi_2 (\rho_2)  &= 
\begin{pmatrix}
0.509	&	 0.257 + 0.329 i \\
0.257 - 0.329 i  	& 	0.491
\end{pmatrix}\,.
\end{align}

\begin{figure*}[t!]
\begin{minipage}[t]{1\textwidth}
\centering
  \includegraphics[width=0.95\linewidth]{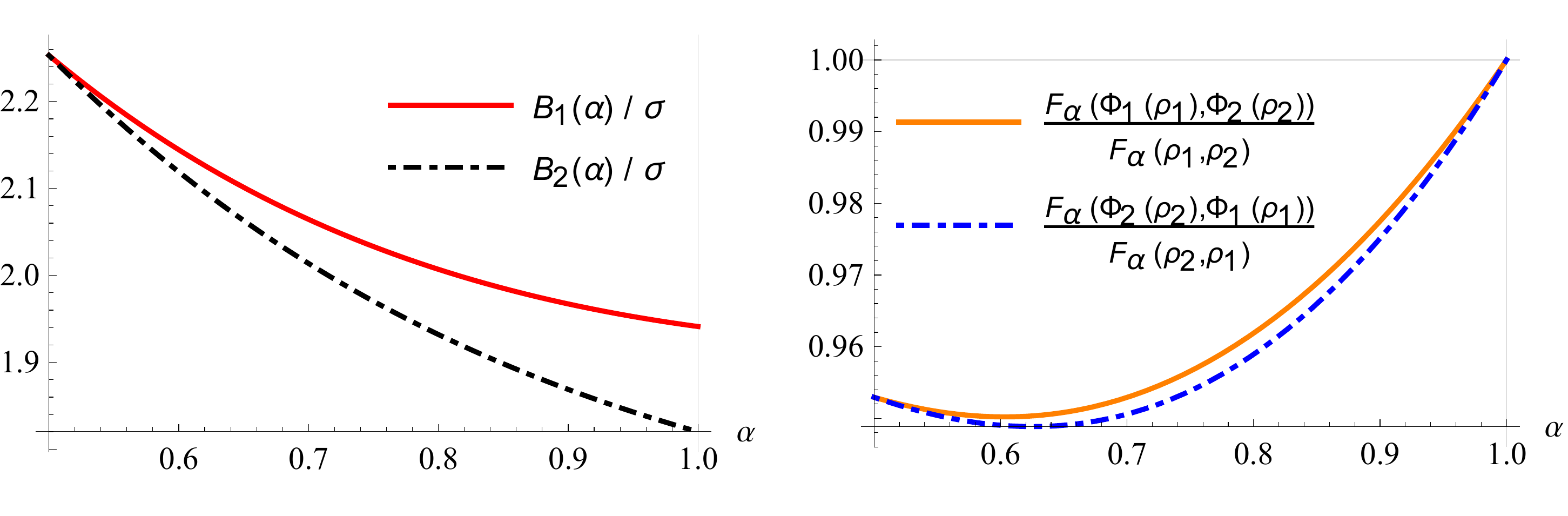}
  \caption{The experimentally determined upper bounds of the unknown parameter $\sigma$ (color online). Here, the coupling is a 5 mm quartz plate. The panel on the left-hand side illustrates the upper bounds of the unknown $\sigma$. We see that as $\alpha$ approaches 1, this measurement gives the tighter upper bound $B_2$ as $1.82 \times \sigma$, where the actual standard deviation is $\sigma = 5.68 \times 10^{11}$ Hz. The panel on the right-hand side presents the fractions $F_\alpha \big( \Phi_1 (\rho_1), \Phi_2 (\rho_2) \big)/F_\alpha (\rho_1,\rho_2)$ and  $F_\alpha \big( \Phi_2 (\rho_2), \Phi_1 (\rho_1) \big)/F_\alpha (\rho_2,\rho_1)$, corresponding to the bounds in the left panel. We see, that in both cases the fraction is less than one for $\alpha \in [1/2,1)$, and thus the use of the bounds $B_1(\alpha)$ and $B_2(\alpha)$ is justified. }
  \label{measurementdata5mm_dummy}
\end{minipage}
\end{figure*}

In figure \ref{measurementdata5mm_dummy}, we plot the upper bounds of Eq.~\eqref{ourbound1} and \eqref{ourbound2} for the unknown $\sigma$ as a function of the parameter $\alpha$. The upper bounds are given in units of the actual value of standard deviation $\sigma = 5.68
 \times 10^{11}$ Hz (about 1.24 
 nm in wavelength). We note that the freedom to choose $\alpha$ leads to significantly tighter bounds: increasing $\alpha$ towards $\alpha = 1$ tightens the upper bound significantly from $\tilde{\sigma} \leq 2.22 \sigma$ to $\tilde{\sigma} \leq 1.82 \sigma$. In this case, it is clear that the the lack of input symmetry in $F_\alpha$ leads to two different bounds $B_1(\alpha)$ and $B_2(\alpha)$, and $B_2(\alpha)$ leads to a tighter upper bound. 

\begin{figure}[t]
\centering
  \includegraphics[width=0.95\linewidth]{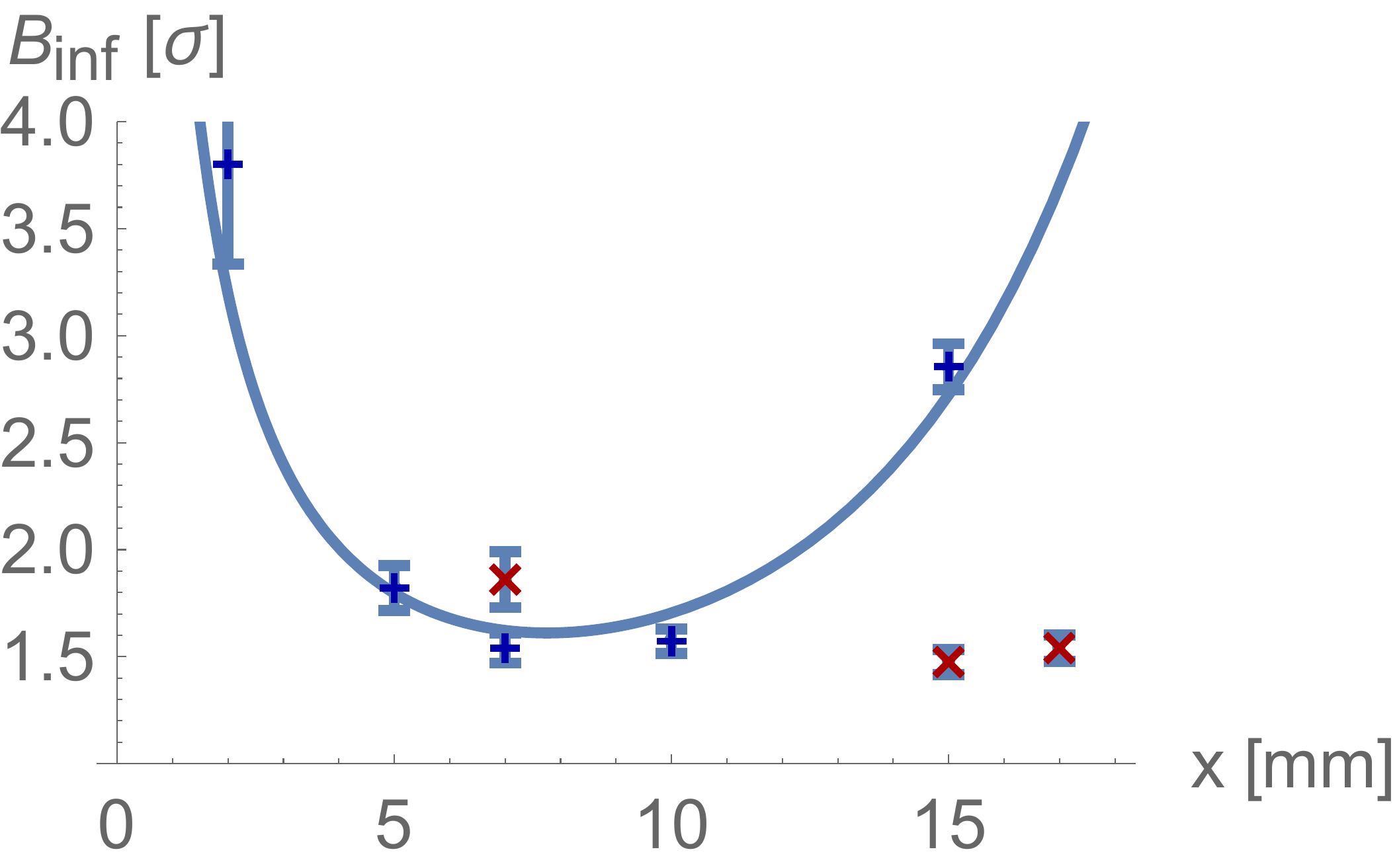}
  \caption{The experimentally determined tightest upper bounds of the unknown parameter $\sigma$ for different couplings $U$ (color online). The $x$ axis is the used quartz plate combination thickness, corresponding to different couplings and $B_{inf}$ is the smallest value of the functions $B_1(\alpha)$ and $B_2(\alpha)$ for the optimal value of  $\alpha$, in units of the actual $\sigma$. The blue crosses correspond to measurements where all quartz plates are aligned in the same direction and the blue line is the corresponding theoretical prediction for the measured upper bound as a function of the QP thickness. The slanted red crosses correspond to measurements where the orientation of each quartz plate was fixed by rotating it in randomly chosen angle. The slanted red cross at 7 mm corresponds to combination of a 2 mm and a 5 mm quartz plates, the slanted red cross at 15 mm corresponds to a combination of a 5 mm and a 10 mm quartz plate, and the slanted red cross at 17 mm corresponds to a combination of a 2 mm, a 5 mm and a 10 mm quartz plate, all rotated in randomly picked and unknown angles. The error bars are due to the photon counting statistics, and they are standard deviations of bound values calculated by the Monte-Carlo method.}
  \label{all_bounds}
\end{figure}

The protocol was repeated for multiple other couplings, implemented with different thicknesses $x$ of quartz plate combinations rotated in different orientations. The results are summarized in Fig.~\ref{all_bounds}
. Here, the thicknesses $x$ are listed for completeness, but knowledge of the QP thickness - as any other properties of the coupling - is irrelevant for our protocol. For each coupling, the validity of the bounds was checked and the tightest bound was determined by comparing $B_1(\alpha)$ and $B_2(\alpha)$ for different values of $\alpha$, as explained for the example case of 5 mm quartz plate in Fig.~\ref{measurementdata5mm_dummy}. 

In figure \ref{all_bounds}, the blue crosses are the tightest upper bounds determined with the quartz plate combinations of different thickness $x$ with the optical axes in the same direction. The solid blue line is the corresponding theoretical prediction, plotted by using the solution of polarization dynamics in \cite{liu2011}. We note that first the bounds become tighter as the thickness increases, but after 7 mm the bounds  become less and less tight. This tells us that from 2 to 7 mm, the evolved probe states $\Phi_1(\rho_1)$ and $\Phi_2(\rho_2)$ become less and less similar to each other w.r.t.~$\alpha$-fidelity, decreasing $F_{\alpha} \big(\Phi_1(\rho_1), \Phi_2(\rho_2)\big)$ as the thickness $x$ increases. After 7 mm, the bounds become less and less tight, which means that the evolved states $\Phi_1(\rho_1)$ and $\Phi_2(\rho_2)$ become more and more similar, increasing $F_{\alpha} \big(\Phi_1(\rho_1), \Phi_2(\rho_2)\big)$. The measurement with coupling that corresponds to 20 mm quartz plate combination is not shown in Fig.~\ref{all_bounds}
, because in that case the evolved probe states became more similar than the initial states, and thus we got $F_{\alpha} \big( \rho_1, \rho_2\big) < F_{\alpha} \big(\Phi_1(\rho_1), \Phi_2(\rho_2)\big)$, which meant that the measurement data does not give any non-trivial bounds.

The slanted red crosses in Fig.~\ref{all_bounds} are the tightest upper bounds determined with the quartz plate combinations of different thickness $x$ with the optical axes fixed in randomly chosen and unknown directions, corresponding to truly unknown couplings. 
For these cases, the orientation of quartz plates in the combinations is set by choosing random and fixed direction independently for each of the quartz plates. Then, state tomography is performed for the evolved polarization states after interaction with the frequency in the QP combination. 
We note that in the case of $x$ = 7 mm the coupling with randomly fixed QP orientations leads to a slightly less tight bound than in the case of the fixed orientation. On the other hand, when $x$ = 15 mm the bounds given by the randomly oriented quartz plates gives significantly tigher bound than the same combination with all the plates set in the same orientation. This case is the tightest bound we achieved in all of the measurements. Thus, with our protocol modifications of the system-probe coupling can be used to tighten the obtained bounds even if the effect of the modification in the resulting probe dynamics could not be analyzed.

In \cite{muller-lennert} it was shown that the quantum R\'enyi divergences are continuous functions w.r.t.~their argument states. As the $\alpha$-fidelities are continuous functions of the R\'enyi divergences, also $\alpha$-fidelities are continuous, and thus small deviations in the tomography are not critical for the determined $\alpha$-fidelity values. To experimentally test the sensitivity of our protocol with respect to the precision of probe tomography, we performed initial probe state tomography again by using only 10 second integration time for each basis. Smaller sample size changed the resulting states only slightly, and the average difference between the bounds obtained from 60 s and 10 s tomography was 4.19 \% of the bound with 60 s tomography. This serves as experimental evidence for the robustness of our approach.

As illustrated by Eq.~\eqref{ourbound1}, Eq.~\eqref{ourbound2}, and our experimental data, this approach can give only analytically derived bounds for the unknown parameter to be determined and not its actual value. This is a trade-off of allowing the measurement protocol to function with no knowledge of the coupling $U$ which the probing is based on.

\section{Conclusions and Outlook}\label{conclusions}

In this paper we presented the first experimental realization of a quantum probing protocol that works with unknown system-probe coupling. Our system of interest was the frequency of a photon and the unknown parameter of interest was the standard deviation of its Gaussian spectrum. The frequency spectrum was realized in two Gaussian distributions with different means and each of them were coupled to their corresponding polarization probes. Comparison of the evolved polarization states was applied in the generalized data processing inequality which gave us analytically derived upper bounds for the standard deviation as functions of the measured probe states.

We repeated the measurement for multiple system-probe couplings, implemented with quartz plate combinations of different thicknesses. To emphasize how our protocol does not rely on any knowledge of the coupling, we performed it by using also combinations of quartz plates whose orientations were fixed in randomly chosen and unknown angles. The experimentally determined standard deviation upper bounds showed that there was no benefit of knowing the quartz plate orientations and actually the tightest upper bound for the unknown parameter was obtained with an unknown coupling. These experiments pave the way for new kind of measurement protocols which do not rely on high precision implementation and control of some desired coupling – or knowing the coupling scheme at all. On a more foundational level, these results broaden the understanding of the limitations and possibilities of measurements more generally. 

In this work we have performed full tomography for both evolved probe systems. For qubit systems tomography is fast in many cases, but for higher dimensional systems the number of parameters to be determined increases quadratically as a function of the dimension. As our probing protocol requires only the values of $\alpha$-fidelities the natural question arises: is it necessary to perform full tomography for both probes?\\
If tomography of the first measured probe system shows that its density matrix has full rank, full tomography of the other probe system is necessary. 
This is a consequence of a result in \cite{teiko}, which states that the fidelity $F_{1/2}(\sigma,\xi)$ between a known fixed reference state $\sigma$ and an entirely unknown state $\xi$ can be estimated without full tomography if and only if $\sigma$ does not have full rank. This result is easily generalized to $\alpha$-fidelities. Thus, if tomography is done first for one of the evolved probes and its density matrix is not full rank, the $\alpha$-fidelity can be measured without full tomography of the second probe. The mathematical tools for constructing such measurements have been introduced in \cite{teiko} and experimentally implemented to fidelity of two-photon polarization states in \cite{lyyra}. This demonstrates that estimating $\alpha$-fidelities for our probing purposes can be performed more efficiently, especially in the case for higher dimensional polarization probes. Furthermore, If the first evolved probe state is determined as pure, namely $\Phi_1 (\rho_1) = \ket{\phi}\bra{\phi}$, full tomography for the other evolved probe state is not necessary. In this case we have $F_\alpha (\Phi_1 (\rho_1),\Phi_2 (\rho_2)) = \left( \bra{\phi} \Phi_2 (\rho_2) \ket{\phi} \right)^\alpha$ , where $\bra{\phi} \Phi_2 (\rho_2) \ket{\phi}$ is the probability of outcome $\ket{\phi}\bra{\phi}$ in the binary projective measurement $\{\ket{\phi}\bra{\phi},\,\textbf{1} - \ket{\phi}\bra{\phi}\}$.\\ 
Also a priori information about the channels can be exploited: If some matrix elements are known to remain invariant in channel $\Phi_1$ and/or $\Phi_2$, it suffices to measure only the matrix elements that change. In our case the diagonal elements remain almost unchanged and thus the number of projective measurement bases could be reduced from 3 to 2, but for higher dimensional probe systems the improvement could be more significant.

In the polarization-frequency model, the transition from Markovian to non-Markovian polarization dynamics has been detected experimentally \cite{liu2011}. In that case, the relation of the heights of two Gaussian peaks in a double peaked frequency spectrum controlled the transition when the widths and the distance of the Gaussians were fixed. With some modifications, our approach could be used to estimate the relative heights, and possibly to deduce also the Markovian or non-Markovian character of the dynamics. This would allow making accurate conclusions about the global properties of the polarization dynamics by performing measurements only at single unknown point in time.

\section{Acknowledgements}
HL acknowledges the financial support from the University of Turku Graduate School (UTUGS) and the Laboratory of Quantum Optics of University of Turku and the useful discussions with Sina Hamedani Raja and Mikko Tukiainen. OS acknowledges the financial support from Magnus Ehrnrooth Foundation. SB thanks Turku Centre for Quantum Physics, Department of Physics and Astronomy, University of Turku for their warm hospitality during his visit when this work was initiated. SB also acknowledges support from Interdisciplinary Cyber Physical Systems (ICPS) programme of the Department of Science and Technology (DST), India, Grant No.:DST/ICPS/QuST/Theme-1/2019/6. This work was financially supported by the Academy of Finland via the Centre of Excellence program (Project no. 312058).

\end{document}